\documentclass[fleqn,10pt]{wlscirep}
\usepackage{times}

\usepackage{accents}
\usepackage{mathtools}
\usepackage{soul}
\usepackage{color}
\usepackage{cancel}
\usepackage{relsize}
\def\sm{\textbf{\tiny m}}
\def\sn{\textbf{\tiny n}}

\title{The common origin of symmetry and structure in genetic sequences}

\author[1,*]{Giampaolo Cristadoro}
\author[2]{Mirko Degli Esposti}
\author[3]{Eduardo G. Altmann}
\affil[1]{Dipartimento di Matematica e Applicazioni, Universit\`a di Milano-Bicocca, 20125 Milano, Italy}
\affil[2]{Dipartimento di Informatica, Universit\`a di Bologna, 40126 Bologna, Italy}
\affil[3]{School of Mathematics and Statistics, University of Sydney, 2006 NSW, Australia}
\affil[*]{giampaolo.cristadoro@unimib.it}

\begin{abstract}
  Biologists have long sought a way to explain how statistical properties of  genetic sequences emerged and are maintained through evolution. On the one hand, non-random structures at different scales indicate a complex genome organisation. On the other hand, single-strand symmetry has been scrutinised using neutral models in which correlations are not considered or irrelevant, contrary to empirical evidence.
  Different studies investigated these two statistical features separately, reaching minimal consensus despite sustained efforts. Here we unravel previously unknown symmetries in genetic sequences, which are organized hierarchically through scales in which non-random structures are known to be present. These observations are confirmed through the statistical analysis of the human genome and explained through a simple domain model. These results suggest that domain models which account for the cumulative action of mobile elements can explain simultaneously non-random structures and symmetries in genetic sequences.

\end{abstract}
\begin{document}

\flushbottom
\maketitle
\thispagestyle{empty}

\section*{Introduction}

Compositional  inhomogeneity at different scales has been  observed in DNA since the early discoveries of 
long-range spatial correlations,  pointing to a complex organisation of genome sequences  \cite{PBGHSSS92,LK92,V92}. 
While the mechanisms responsible for these observations have been intensively debated \cite{A92,N92, P92, KB93, LMK94, BS12}, several investigations indicate the patchiness and mosaic-type domains of DNA as playing a key role in the existence of large-scale {\it structures} \cite{ KB93, PBHSSG93, BGRO96}.
Another well-established statistical observation is the {\it symmetry} known as ``Second Chargaff Parity Rule" \cite{RKC68}, which appears universally over almost all extant genomes  \cite{R91,MB06,NA06}. In its simplest form, it states that on a single strand the frequency  of a nucleotide is approximately equal to the frequency of its complement \cite{FTW92,P93, BF99, QC01, PHB02}. This original formulation has been later extended  to the frequency of  short ($n \simeq 10$)  oligonucleotides and their reverse-complement  \cite{PHB02, KFCHZZL09, ABGRPF13}. While the first Chargaff parity rule \cite{C51} (valid in the double strand) was instrumental for the discovery of the double-helix structure of the DNA, of which it is now a trivial consequence, the second Chargaff parity rule remains of mysterious origin and of uncertain functional role.  Different mechanisms that attempt to explain its origin have been proposed during the last decades \cite{BF99,BF99b,LL99,B06,ZH10}.  Among them, an elegant  explanation \cite{B06,SCRH16} proposes that strand symmetry arises from the repetitive action of transposable elements.

Structure and symmetry are in essence two  independent observations:  Chargaff symmetry in the frequency of short oligonucleotides ($n \simeq 10$) does not rely on the actual positions of the oligonucleotides in the DNA, while correlations depend on the ordering and are reported to be statistically significant even at large distances (thousands of bases). Therefore, the mechanism shaping the complex organization of genome sequences could be, in principle, different and independent from the mechanism enforcing symmetry. 
However, the proposal of transposable elements \cite{M84, F12} as being a key biological processes in both cases suggests that these elements could be the vector of a deeper connection.

In this  paper  we start with a review of known results on statistical symmetries of genetic sequences and proceed to a detailed analysis of the set of chromosomes of Homo Sapiens. Our main empirical findings are: (i) Chargaff parity rule extends beyond the frequencies of short oligonucleotides (remaining valid on scales where non-trivial structure is present); and (ii) Chargaff is not the only symmetry  present in  genetic sequences as a whole and there exists a hierarchy of symmetries nested at different structural scales. We then propose a model to explain these observations. The key ingredient of our model is the reverse-complement symmetry for domain types, a property that can be related to the action of transposable elements indiscriminately on both DNA strands. Domain models have been used to explain  structures (e.g., the patchiness and long-range correlations in DNA), the significance of our results is that it indicates that the same biological processes leading to domains can explain also the origin of symmetries observed in the DNA sequence.

\section*{Results}
\subsection*{Statistical Analysis of Genetic Sequences}
We  explore  statistical properties of genetic  sequences  $\textbf{s}= \alpha_1 \alpha_2 \cdots  \alpha_{N}$,  with $\alpha_i \in \{A,C,T,G\}$, by quantifying the frequency of appearance in  $\textbf{s}$  of a given pattern of symbols (an observable $X$). For instance, we may be interest in the frequency of the codon ACT in a given chromosome. More generally, we count the number of times a given symbol $\alpha_0$ is separated from another symbol $\alpha_1$ by a distance $\tau_1$, and this from $\alpha_2 $ by a distance $\tau_2$, and so on. The case of ACT corresponds to $\alpha_0=A, \alpha_1=C, \alpha_2=T, \tau_1=1, \tau_2=1$.
We denote  $\underaccent{\bar}{\alpha}:= (\alpha_0,  \alpha_1, \cdots, \alpha_k)$ a selected finite sequence of symbols, and by $\underaccent{\bar}{\tau} := (\tau_1, \cdots, \tau_k )$ a sequence of gaps. For shortness, we denote this  couple  by $X := ( \underaccent{\bar}{\alpha} , \underaccent{\bar}{\tau} )$ and the {\it size} of the observable $X$ by $\ell_X=\sum_{i} \tau_i+1$.
The frequency of occurrence of an observable $X$  in the sequence $\textbf{s}$ is obtained counting how often it appears varying the starting point $i$ in the sequence:
\begin{equation}\label{def.statisticsX}
P(X):=\frac1{N'}  \#_i  \left\{ s_i=\alpha_0, \,  s_{i+\ell_1}=\alpha_1,\,  \cdots , 
s_{i+\ell_k}=\alpha_k   \right\},\qquad \ell_j=\sum_{r=1}^{j}\tau_r
\end{equation}
where $N'=N-\ell_X+1$. As a simple example, for the choice of   $X=\left( (A,C,G), (1,2) \right)$  in the sequence $\textbf{s}=GGACCGGCCACAGGAA$ we have $N=16$, $N'=13$, and $P(X)=2/13$.
All major statistical quantities numerically investigated in literature can be expressed in this form, as we will recall momentarily.  

The main advantage of the more general formulation presented above is that it allows to inspect both the  role of  symmetry (varying $\underaccent{\bar}{\alpha}$) and structure (varying scale separations $\underaccent{\bar}{\tau}$)  and it thus permits a systematic exploration of their interplay.  We say that a sequence has the symmetry $S$ at the scale $\ell$ if for any observable $X$ with length $\ell_X=\ell$ we have, in the limit of infinitely long $\textbf{s}$,
\begin{equation}\label{eq.S}
P(X) = P(S(X))
\end{equation}
where $S(X)$ is the observable symmetric to $X$.

We start our exploration of different symmetries $S$ with a natural extension to observables $X$ of the reverse-complement symmetry considered by Chargaff. 
The reverse complement of an oligonucleotide $\alpha_1\alpha_2,\ldots, \alpha_n$ of size $n$ is $\hat{\alpha}_n\hat{\alpha}_{n-1}\ldots\hat{\alpha}_1$, where $\hat{A}=T, \hat{T}=A, {\hat{C}=G,\hat{G}=C}$ (e.g., the reverse-complement of $CGT$ is $ACG$). For our more general case it is thus natural to consider that the observable symmetric to
\begin{equation*}
  X  = \Big( (\alpha_0,  \alpha_1 \cdots, \alpha_k), (\tau_1, \tau_2 \cdots, \tau_k )\Big)
  \end{equation*}
is 
\begin{equation}\label{EC}
 \hat{X}:=\Big(  (\hat{\alpha}_k, \hat{ \alpha}_{k-1} \cdots, \hat{\alpha}_0 ) , ( \tau_k, \tau_{k-1} \cdots, \tau_1) \Big).
\end{equation}
This motivates us to conjecture the validity of an extended Chargaff symmetry
\begin{equation}\label{eq.extended}
P(X) = P(\hat{X}).
\end{equation}
This is an {\it extension} of Chargaff's second parity rule because $X$  may in principle be an observable involving (a large number of) distant nucleotides and thus equation~ (\ref{eq.extended})  symmetrically connects  structures  even at large scales. One of the goals of our manuscript is to investigate the validity of Eq.~(\ref{eq.extended}) at different scales, which will be done by choosing observables $X$ of size $\ell_X$ of up to millions of base pairs. 

By combining $P(X)$ of different observables $X$ this extended Chargaff symmetry applies to the main statistical analyses already investigated in literature, unifying numerous previously unrelated observations of symmetries. As paradigmatic examples we have:
\begin{itemize} 
\item[-] the \emph{frequency of a given oligonucleotide} $\underaccent{\bar}{\omega}=\omega_1 \omega_2 \cdots \omega_k$  can be computed as $P(X)$ with the choice $\underaccent{\bar}{\alpha}=(\omega_1, \omega_2, \cdots, \omega_k)$ and $\tau=(1,1,\cdots 1)$. Equation~(\ref{eq.extended}) implies that  the frequency of an oligonucleotide is equal to the frequency of its reverse-complement symmetric  and thus implies the second Chargaff parity rule, a feature that has been extensively confirmed\cite{PHB02, KFCHZZL09, ABGRPF13} to be valid for short oligonucleotides $\ell_X \le 10$. We report few examples of frequencies of dinucleotides ($\ell_X = 2$) in human chromosome $1$:  
$P(AG)= 7.14\% \approx  P(CT)= 7.13\% \ne P(GA)= 6.01 \% \approx P(TC)= 6.01\%$, in agreement with symmetry~(\ref{eq.extended}). Note that, the validity of Second Chargaff Parity rule at small scales ($\ell_X=2$ for dinucleotides) is not enough to enforce  equation (\ref{eq.extended}) for generic observables $X$ (e.g., of size $\ell_X \gg 100$);
\item[-]
the \emph{autocorrelation function $C_{\omega}(t)$} of nucleotide $\omega$ at delay $t$ is the central quantity in the study of long-range correlations in the DNA. It corresponds to the choice $\underaccent{\bar}{\alpha}=(\omega, \omega)$ and $\underaccent{\bar}{\tau}=(t)$.  Equation~(\ref{eq.extended}) predicts the symmetry $C_{\omega}(t)=C_{\hat{\omega}}(t)$.  In the specific  case of dinucleotides, such relation has been  remarked in Ref.~\cite{Li97}. Our result holds for any oligonucleotide $\omega$;
\item [-]
  the \emph{recurrence-time distribution $R_{\omega}(t)$ }of  the first return-time  between two consecutive appearances of the oligonucleotide $\omega$ is studied in Refs.~\cite{ABPGF09, FS12}. By using elementary arithmetic and common combinatorial techniques, it is easy to see that $R_{\omega}(t)$  can be  in fact written as a sum of different $P(X)$.  Equation~(\ref{eq.extended}) hence predicts $R_\omega(t) = R_{\hat{\omega}}(t)$. 
  This symmetry was observed for  oligonucleotides in Ref.~\cite{TPSRBFA17}.
\end{itemize} 

This brief review of previous results shows the benefits of our more general view of Chargaff's second parity rule and motivates a more careful investigation of the validity of different symmetries at different scales $\ell$.

\subsection*{Symmetry and structure in Homo Sapiens}
 We  now investigate the existence of new symmetries in the human genome. In order to  disentangle the role of different symmetries at different scales~$\ell$ we construct a family of observables $X=(\underaccent{\bar}{\alpha},\underaccent{\bar}{\tau})$ for which we can scan different length scales by varying the gaps vector $\underaccent{\bar}{\tau}$. Particularly useful is to fix all gaps in $\underaccent{\bar}{\tau}$ but a chosen one $\tau_j$, and  let it vary through different scales.  To  be more specific consider the following construction: 
given two patterns $X_A=(\underaccent{\bar}{\alpha}_A, \underaccent{\bar}{\tau}_A)$ and $X_B=(\underaccent{\bar}{\alpha}_B, \underaccent{\bar}{\tau}_B)$ we look for their  appearance in a sequence, separated by a  distance $\ell$. This is equivalent to look for
 composite observable $Y= ((\underaccent{\bar}{\alpha}_A , \underaccent{\bar}{\alpha}_B), (\underaccent{\bar}{\tau}_A\, \ell \, \underaccent{\bar}{\tau}_B))$ or, for simplicity, $Y=:(X_A,X_B; \ell)$.   
 We consider two patterns $X_A,X_B$  of small (fixed) size $\ell_{X_A}, \ell_{X_B}$ and  we vary their separation $\ell$ to investigate the change in the role of different symmetries.

 To keep the analysis feasible, we scrutinize the case where $X_A$ and $X_B$ are dinucleotides separated by a distance $\ell$ from each other. This goes much beyond the analysis of the frequencies of short oligonucleotides mentioned above  because with $\ell$ ranging from $1$ to $10^7$ we span ranges of interests for structure and long-range correlations. This choice has two advantages: by keeping the number of nucleotides in  each $X$ small we improve statistics, but at the same time we still differentiate  $\hat{X}$ from the simpler complement transformation. 

\begin{figure}[h] 
\centering
\includegraphics[width=0.7\columnwidth]{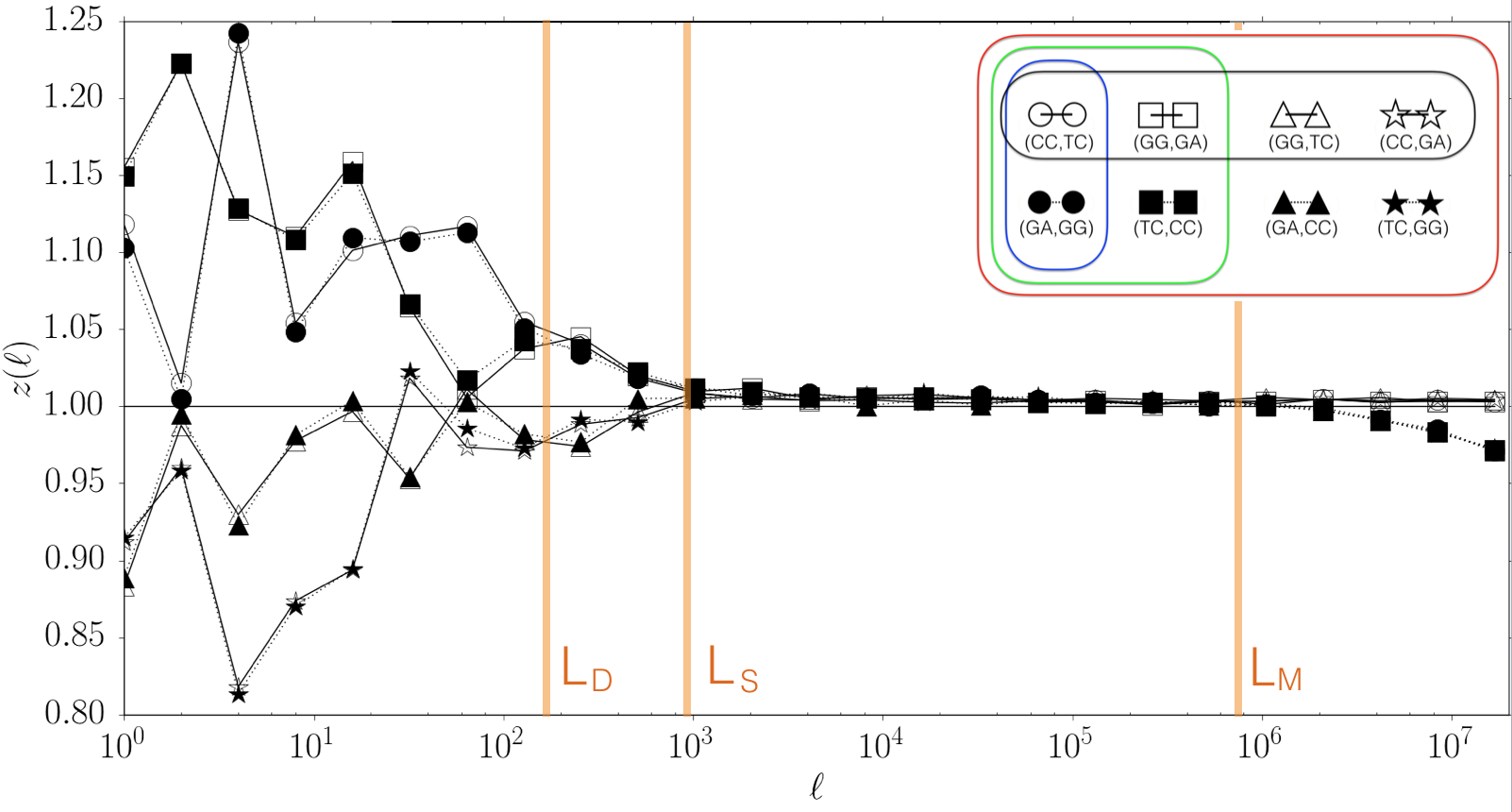}
\caption{\textbf{Symmetrically related cross-correlations in Homo Sapiens - Chromosome 1.} The normalized cross-correlations $z_{[CC,TC]}(\ell)$ as a function of the scale $\ell$, together with that of  its symmetrical related companions.    Symmetries are significant also at scales where non trivial correlations are present $z\neq1$.}
\label{fig.1}
\end{figure}

In order to compare results for pairs $X_A,X_B$ with different abundance,  we  normalise our observable by the expectation of independence appearance of $X_A,X_B$ obtaining
\begin{equation}\label{eq.z}
z_{[X_A,X_B]}(\ell)=\frac{P(X_A,X_B; \ell)}{P(X_A)P(X_B)}.
\end{equation}
Deviations from $z=1$ are  signatures of structure (correlations). Table~\ref{tab.1} shows the results for chromosome 1 of  Homo Sapiens, using a representative set of eight symmetrically related  pairs of dinucleotides at a small scale $\ell = 4$. The results show that $z$ is significantly different from one and that Chargaff symmetric observables $(Y,\hat{Y})$ appear with similar frequency, in agreement with conjecture~(\ref{eq.extended}).  Figure~\ref{fig.1} shows the same results of the Table varying logarithmically the scale $\ell$ from   $\ell=1$ up to  $\ell \simeq 10^7$ ( more precisely we use $\ell =2^i$ with $i\in \{0,1,2,..,24\}$). At different scales $\ell$ we see that a number of lines (observables $X_A,X_B$) coincide with each other, reflecting the existence of different types of symmetries.
\begin{table*}
\begin{center}
\begin{tabular}{|c  | c | c |}
  \hline
 {$X=X_A\;****\;X_B$} & {$P(X)$} & {$z_{[X_A,X_B]}(\ell=4)$} \\ \hline\hline
  CC  **** TC & 0.00401 & 1.236 \\
  GA  ****  GG & 0.00404 & 1.242  \\\hline 
  GG  ****  GA & 0.00366 & 1.127  \\
  TC  ****  CC & 0.00366 & 1.128  \\\hline
  GG  ****  TC & 0.00302 & 0.929 \\
  GA  ****  CC & 0.00299 & 0.922   \\\hline
  CC  ****  GA & 0.00265 & 0.818  \\
  TC  ****  GG & 0.00264 & 0.813  \\\hline
  \end{tabular}
\caption{ \textbf{Chargaff symmetric observables appear with similar frequency in the human chromosome $1$}. Each line contains an observable $X$ constructed combining oligonucleotides $\alpha_1\alpha_2 \ldots \alpha_8$  where $\alpha_1\alpha_2$ equal to $X_A$ , $\alpha_3\alpha_4\alpha_5\alpha_6 $ are arbitrary (any in $\{A,C,T,G\}$), and $\alpha_7\alpha_8$ equal to $X_B$  . Observables related by the extended Chargaff symmetry~(\ref{EC}) appear on top of each other (separated by an horizontal line). The frequency of each observable $P(X)$ 
was computed using Eq.~(\ref{def.statisticsX}) and the normalized version (cross correlation) $z_{[X_A,X_B]}(\ell=4)$ using Eq.~(\ref{eq.z})}. \label{tab.1}
\end{center}
\end{table*}

\begin{figure}[h] 
\centering
\includegraphics[width=0.7\columnwidth]{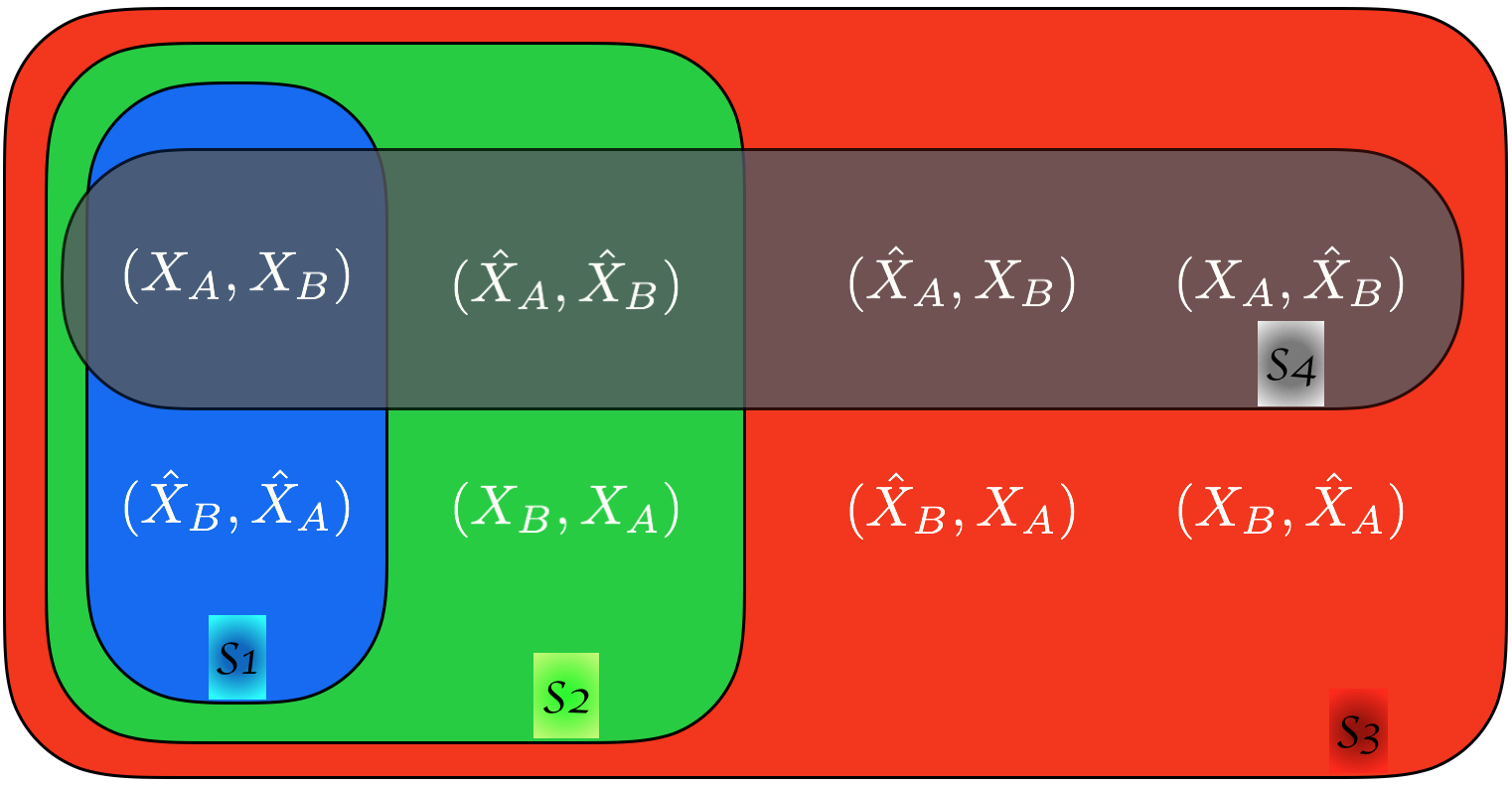}
\caption{\textbf{Sets of symmetrically related observables.}  Starting from a reference observable $Y=(X_A,X_B,\ell)$, the different colors illustrate the nested sets related by symmetries $S1-S4$. The symmetries shown in the figure correspond to: $S1$ (blue, obtained from $\{CRC\}$), $S2$ (green, obtained from $\{CRC,R\}$), $S3$ (red, obtained from $\{R,C\}$),  and $S4$ (black, obtained from $\{RCR,C\}$).}
\label{fig2}
\end{figure}

In order to understand the observations reported above it is necessary to formalize the symmetries that arise as composition of basic transformations. Starting from a reference observable $Y=(X_A,X_B;\ell)$, these symmetries are defined as compositions of the following two transformations:
  \begin{itemize}
  \item[($R$)] reverses the order in the pair:  $ (X_A, X_B; \ell)  \xrightarrow{R} (X_B,X_A; \ell)$.
  \item[($C$)] applies our extended symmetry equation~(\ref{EC})  to the first of the two observable in the pair: $(X_A,X_B; \ell)  \xrightarrow{C}  (\hat{X}_A,{X}_B; \ell)$.
  \end{itemize}
    Note that $RC \neq CR$ (i.e. $R$ and $C$ do not commute), $RR = CC = Id$ (i.e. $R,C$ are involutions), and $CRC$  is the symmetry equivalent to equation~(\ref{EC}). A symmetry $S$ is defined by a set of different compositions of $C$ and $R$. We denote by $\mathcal{S}_S(Y)$  the set of observables obtained applying to observable $Y$ all combinations of transformations in  $S$. For example if $S1=\{CRC\}$ then $\mathcal{S}_{S1}(X_A, X_B; \ell) =\{ (X_A, X_B; \ell), (\hat{X}_B,\hat{X}_A,\ell)  \}$; if $S2=\{CRC,R\}$ then in addition to the set $\mathcal{S}_{S1}$ obtained from $CRC$ we should add the observables obtained by applying $R$ to every element of $\mathcal{S}_{S1}$, that are $R((X_A, X_B; \ell))=(X_B, X_A; \ell)$ and $R( (\hat{X}_B,\hat{X}_A,\ell))=(\hat{X}_A,\hat{X}_B,\ell) $ thus obtaining $\mathcal{S}_{S2}(X_A, X_B; \ell) =\{ (X_A, X_B; \ell), (\hat{X}_B,\hat{X}_A,\ell), (X_B, X_A; \ell),   (\hat{X}_A,\hat{X}_B,\ell)  \} $.
The four symmetries we consider here are shown in Fig.~\ref{fig2}  and correspond to: $S1$ 
(blue, obtained from $\{CRC\}$ and corresponding to the extended Chargaff~(\ref{eq.extended})), $S2$ (green, obtained from $\{CRC,R\}$), $S3$ (red, obtained from $\{R,C\}$),  and $S4$ (black, obtained from $\{RCR,C\}$).
We can now come back to Fig.~\ref{fig.1} and interpret the observations as follows:  at  scales $\ell<L_D$  curves are significantly different from $z=1$ and appear in pairs  (same symbol, symmetry $S1$) which almost coincide even in the seemingly random fluctuations;  around $\ell \simeq  L_D \approx 2\;10^2$  two pairs merge forming two groups of four curves each (symmetry $S2$).  At larger scales $\ell \ge L_S \approx 10^3$ all curves coincide (symmetry $S3$) at $z \approx 1$ (no structure). At  very large scales $\ell > L_M \approx 10^6$ two groups of four observables separate (symmetry $S4$). Similar results are obtained for all choice of dinucleotides and for all chromosomes (see SI: Supplementary data~\cite{zenodo}). These results suggest that: (i) the extended Chargaff symmetry we conjectured in  Eq.~(\ref{eq.extended}) is valid  up to a critical scale $L_M \simeq 10^6$; (ii)  there are other characteristic scales  connected to the  other symmetries.

%
The scale-dependent results discussed above motivate us to quantify the strength of validity of symmetries at different scale $\ell$.  This is done computing for each symmetry $S$ an indicator $I_{S}(\ell)$ that measures the distance between the curves $z(\ell)$ of symmetry-related pairs $(X_A,X_B)$ and compares this distance to the ones that are not related by $S$.  
More precisely, for a given  reference pair $Y_{\textrm{ref}}=(X_A,X_B)$ and  symmetry $S\in \{S1,S2,S3,S4\}$,  we consider the  following distance of $Y_{\textrm{ref}}$ to the set $\mathcal{S}_S$ of observables obtained from symmetry $S$:   
\begin{eqnarray}
d_{\ell}^{}(Y_{\textrm{ref}}; S )&:=&\frac1{ |\mathcal{S}_S|} \mathlarger{\sum}_{Y \in \mathcal{S}_S(Y_{\textrm{ref}})}{ \frac{ \left[ z_{[Y]}{(\ell)} - z_{[Y_{\textrm{ref}}]}(\ell)\right]^2 }{\sigma^2(\ell)}} 
\end{eqnarray}
where $\sigma(\ell)$ denotes the standard deviation of $z(\ell)$ over all $Y$. 
We  then average over the set $\mathcal{A}$ of all $Y_{\textrm{ref}}$ (all possible pairs $X_A,X_B$) to obtain a measure of the strength of symmetry $S$ at scale $\ell$ given by 
\begin{equation}
I_{S}(\ell):=\frac1{2|\mathcal{A}|} \mathlarger{\sum}_{Y_{\textrm{ref}}\in \mathcal{A} }{ d_{\ell}(Y_{\textrm{ref}}; \mathcal{S} )}
\end{equation}
Note that  $I_{S}(\ell)=0$ indicates full validity of the symmetry $S$ at the scale $\ell$ ($z$ is the same for all $Y$ in $\mathcal{S}$) and $I_{S}(\ell)=1$ indicate that $S$ is not valid at scale $\ell$ ($z$ varies in $\mathcal{S}$ as much as it varies in the full set).


\begin{figure}[h] 
\centering
\includegraphics[width=0.7\columnwidth]{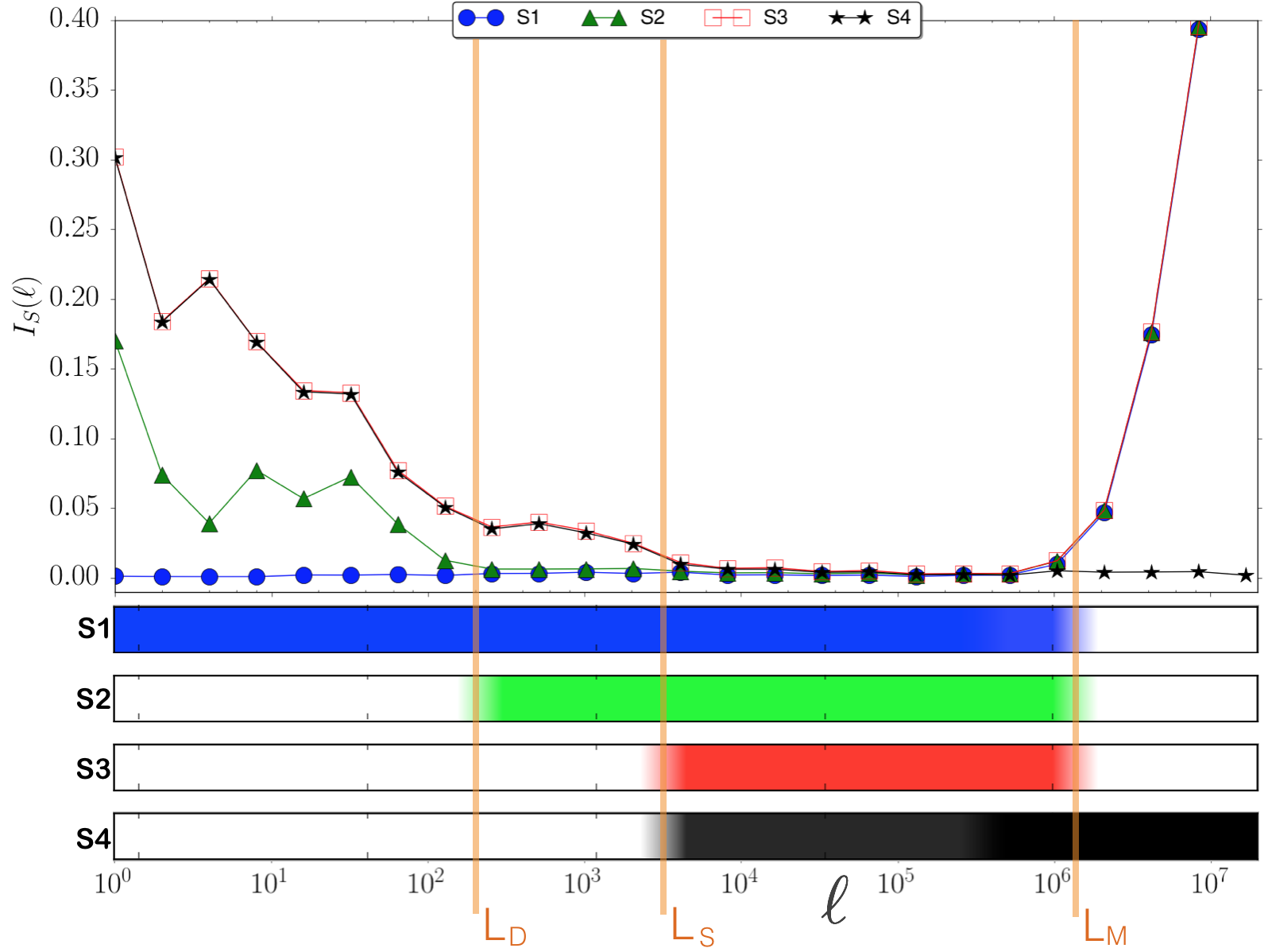}
\caption{\textbf{Hierarchy of symmetries in Homo Sapiens - Chromosome 1.}  [ {\it Upper panel} ] The symmetry index $I_S( \ell)$ as a function of the scale $\ell$, the smaller the value the larger the importance of the symmetry.  
[{\it Bottom panel} ] The color bars helps visualise the onset of the different symmetries:  symmetry is considered present if $0\le I_S \le 0.025$ and   bar is (linearly interpolated) from full color to white, correspondingly.}
\label{fig.3}
\end{figure}
\begin{figure}[h!] 
\centering
\includegraphics[width=0.7\columnwidth]{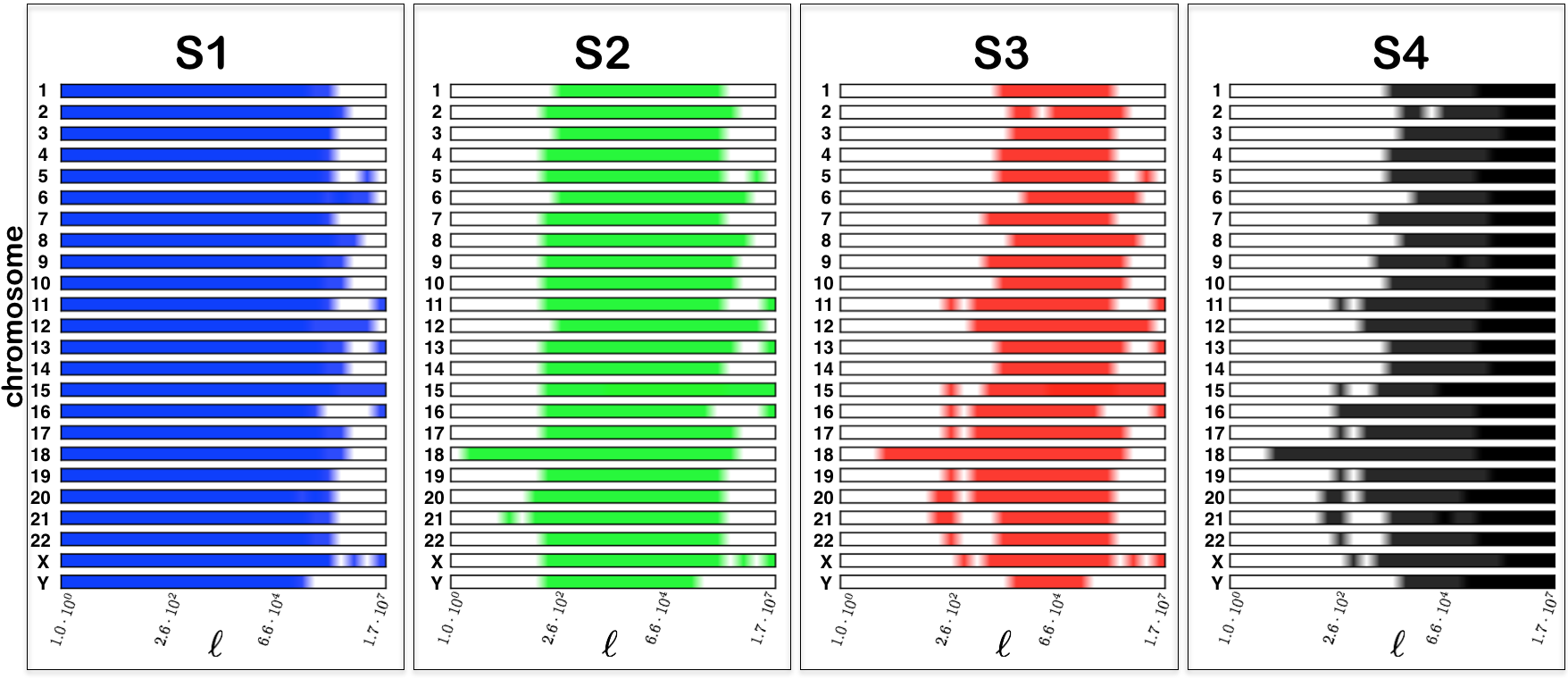}
\caption{\textbf{Hierarchy of symmetries in Homo Sapiens. - All chromosomes.}  The symmetry index $I_S( \ell)$ as a function of the scale $\ell$ for the full set of chromosomes in Homo Sapiens. The brighter the color, the larger is the relevance of the symmetry $S_1,S_2,S_3,$ or $S_4$ (more precisely, if $I_{min}$ is the minimum $I_S$ in each chromosome, $I_S \le 1.05 I_{min}$ is set to full color, $I_S \ge 6.5 I_{min}$ is set to white, with intermediate values interpolated between these extremes.) }
\label{fig.4}
\end{figure}


Figure~\ref{fig.3} shows the results for chromosome 1 and confirms the existence of a hierarchy of symmetries  at different structural scales.  The estimated relevant scales  in chromosome 1 (of total length  $N \approx 2 \times 10^8$)  are $L_D\approx 10^2$, $L_S\approx 10^3$, and $L_M\approx 10^6$.  Note that $L_D$ and $L_M$ are compatible with the known average-size of transposable elements and isochores respectively \cite{BOFZSCMR85, CBCHO07}. Moreover,  the results for all Homo-Sapiens chromosomes, summarised in Fig.\ref{fig.4}, show that not only the hierarchy is present, but also that  the scales $L_D,L_S,$ and $L_M$ are comparable across chromosomes. This remarkable similarity (see also \cite{LSBO98,FZW10, BKB14})  suggests that some of the mechanisms shaping  simultaneously  structure  and symmetry work similarly in every chromosomes and/or act across them (e.g.chromosome rearrangements mediated by  transposable elements). This, and the scales of $L_D, L_S$, and $L_M$, provide a hint on the origin of our observations, which we explore below through the proposal of a minimal model.

\subsection*{A minimal model}

We construct  a minimal domain  model for DNA sequences $\textbf{s}$ that aims to explain the observations reported above. The key ingredient of our model is the reverse-complement symmetry of domain-types, suggested by the fact that transposable elements act on both strands.  Mobile elements are recognised to play a central role in shaping domains and other structures up to the scale of a full chromosome, as well as being considered responsible for the appearance of Chargaff symmetry\cite{B06}. Our model accounts for structures (e.g., the patchiness and long-range correlations in DNA) in a similar way as other domain models do, the novelty is that it shows the consequences to the symmetries of the full DNA sequence.

Motivated by our finding of the three scales $L_D,L_S,$ and $L_M$, our model contains three key ingredients at different length scales $\ell$ (see Fig.\ref{fig.5} for an illustration):
\begin{figure}[ht] 
\centering
\includegraphics[width=0.7\columnwidth]{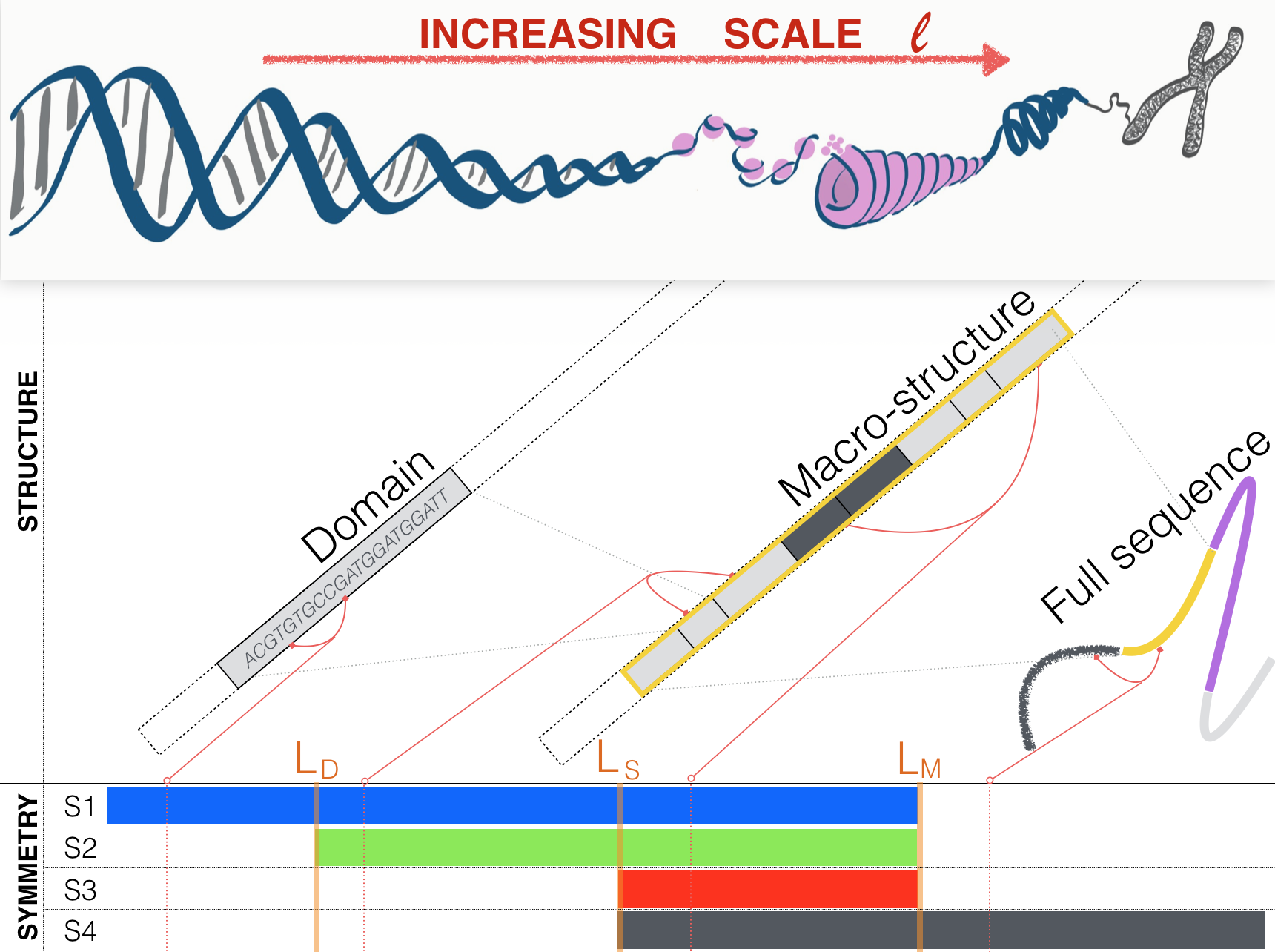}
\caption{\textbf{Structure and symmetry at different scales: domain model.} Structure and symmetry at different scales $\ell$ of genetic sequences can be explained using a simple domain model. Our model considers that the full sequence is composed of macro-structures (of size $L_M$) made by the concatenation of domains (of average size $L_D<L_M$), which are themselves correlated with neighbouring domains (up to a scale $L_D < L_S < L_M$). The biological processes that shapes domains imposes  that,  in  each macrostructure,  the types of domains comes in symmetric pairs. As a consequence, we show that four different symmetries $S_1-S_4$ are relevant at different scales~$\ell$ (see text for details).
}
\label{fig.5}
\end{figure}
\begin{itemize}

\item[(1)] at small scales, a genetic sequence $ \textbf{d}^{} = \alpha_1^{} \alpha_2^{} \cdots  \alpha_{n_{}}^{}$ (of average size $ \langle n_{} \rangle \approx L_D$ ) is generated  as a realization  of a given process $p$. We do not impose a priori restrictions or symmetries on this process. We consider that one realization of this process builds a {\bf domain} of type $p$. For a given domain type, the symmetrically related type is defined by the process $\hat{p}$ as follows:  take a realization ($\alpha_1 \alpha_2 \ldots \alpha_n$) of the process $p$, revert its order $(\alpha_n  \alpha_{n-1}  \ldots \alpha_1)$, and complement each base $(\hat{\alpha}_n \hat{\alpha}_{n-1} \ldots, \hat{\alpha}_1)$, where $\hat{A}=T, \hat{T}=A,\hat{C}=G,\hat{G}=C$;

\item[(2)] at intermediate scales,  a {\bf macro-structure}  is composed as a  concatenation of domains $\textbf{d}_1 \textbf{d}_2 \cdots  \textbf{d}_{m}$ (of average size $\langle m \rangle \approx L_M$),  each domain belonging to one of a few types. We assume that symmetrical related domains (generated by $p$ and $\hat{p}$) appear with the same relative abundance and  size-distribution in a given macro-structure. The concatenation process is done so that domains of the same type tend to form clusters of average size $L_{S}$ such that $L_D<L_{S} \ll L_M$;

\item[(3)] at large scales  $(\gg L_M)$, the full genetic sequence is composed by  concatenations of  macro-structures, each of them governed by different processes and statistics (e.g. different CG content~\cite{PBHSSG93,BGRO96}).
\end{itemize}

 \begin{figure}[ht] 
\centering
\includegraphics[width=0.6\columnwidth]{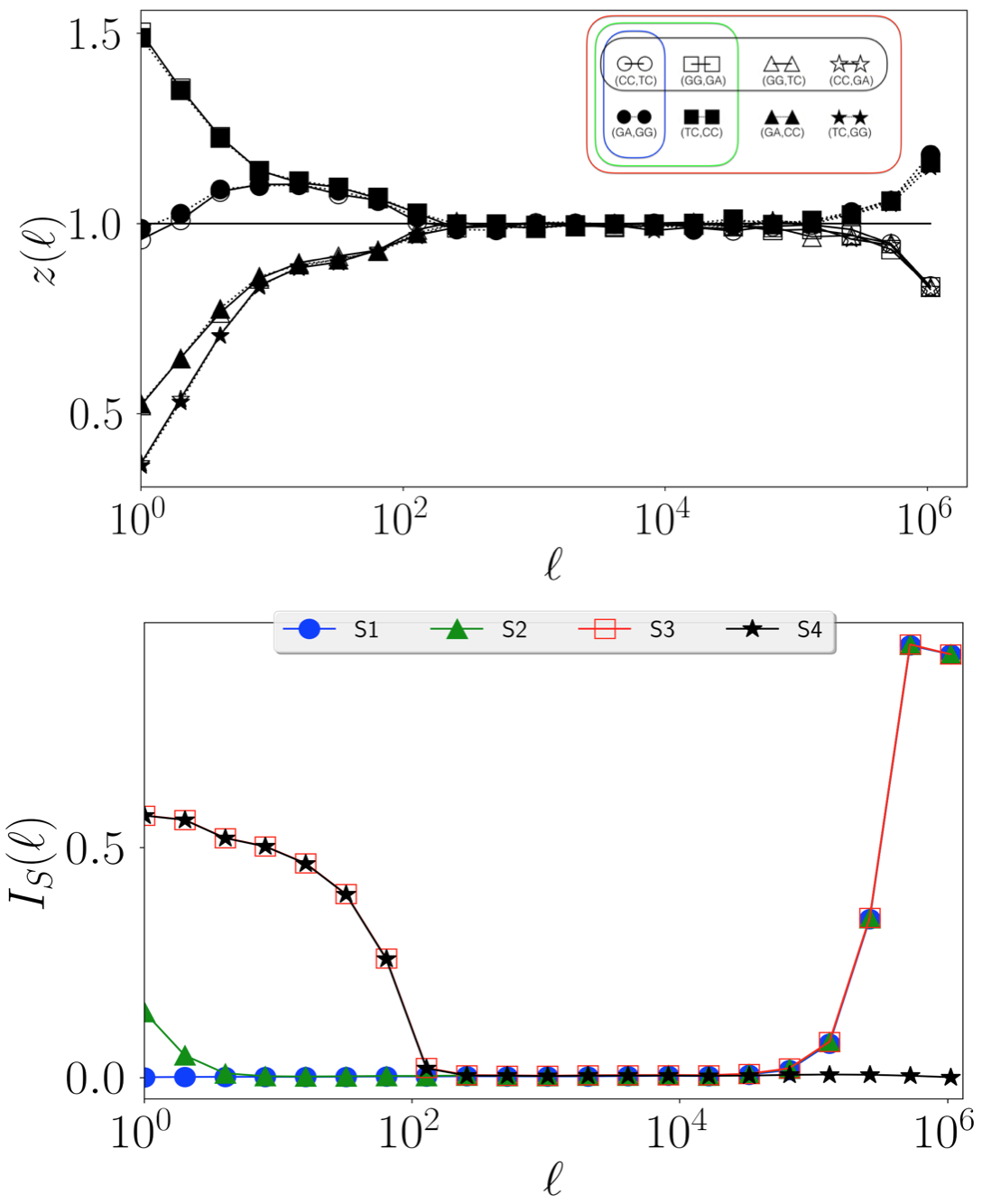}
\caption{ \textbf{Hierarchy of symmetries in a synthetic sequence generated by the domain model}.  The analysis of a synthetic genetic sequence generated by our model reproduces the hierarchy of symmetries observed in the human genome (compare the two panels to Figs.~\ref{fig.1} and~\ref{fig.3}). The synthetic sequence is obtained following steps (1)-(3) of the main text. As  main stochastic processes $p$ we use  Markov chains with invariant probabilities $\mu$ such that $\mu(A) \ne \mu(T) $ and $ \mu(C) \ne \mu(G)$ (no symmetries) (see Materials for details).}
\label{fig.6}
\end{figure}


\subsection*{Statistical properties of the  model and predictions}

We now show how the model proposed above accounts for our empirical observation of a nested hierarchy of four symmetries $S_1$-$S_4$ at different scales. We start generating a synthetic sequence for a particular choice of parameters of the model described above (see  section Methods for details). Figure~\ref{fig.6} shows that such synthetic sequence  reproduces  the same hierarchy of symmetries we detected in Homo Sapiens.

We now argue analytically why these results are expected. The key idea is to note that for different separations $\ell$ (between the two observables $X_A$ and $X_B$) different scales of the model above dominate the counts used to compute $P(X_A,X_B;\ell)$ through Eq.~(\ref{def.statisticsX}) (see Supplementary Information for a more rigorous derivation):

 \begin{itemize}\label{hierarchy}
\item[-]($\ell  \ll L_D$): $P(X_A,X_B,\ell)$ is dominated by $X_A$ and $X_B$  in the same domain. As  domain-types appear symmetrically in each macro-structure, $P(X_A,X_B;\ell)= P(\hat{X}_B,\hat{X}_A;\ell)$. This is compatible with the conjecture~(\ref{eq.extended}).\\ 
\\
$S1=\{CRC\}$\\
$\mathcal{S}_{S1}=\{ ( X_A,X_B;\ell) ,(\hat{X}_B,\hat{X}_A;\ell) \}$.
\\

\item[-] ($L_D \ll \ell \ll L_{S}$): $P(X_A,X_B,\ell)$ is dominated  by $X_A$ and $X_B$ in  different domains. As domains are independent realizations, the  order of $X_A$ and $X_B$ becomes irrelevant and therefore $R$ becomes a relevant symmetry (in addition to $CRC$). If domains of the same type tend to cluster, then  for $L_D < \ell  \ll L_{S}$ the  main  contribution to $P(X_A,X_B,\ell)$ comes from  $X_A$ and $X_B$ in different domains of the {\it same type} (i.e., on different realizations of the same process $p$).\\
\\
$S2 =\{CRC,R \}$\\
$\mathcal{S}_{S2}=\{ ( X_A,X_B;\ell ) ,  (\hat{X}_B,\hat{X}_A;\ell), (X_B,X_A;\ell), (\hat{X}_A,\hat{X}_B;\ell) \}$.

Note that $\mathcal{S}_{S1} \subset \mathcal{S}_{S2}$.
\\

\item[-]($L_{S} \ll \ell \ll L_M$): $P(X_A,X_B,\ell)$ is dominated  by  $X_A$ and $X_B$  in different domains inside the same macro-structure. For $\ell > L_S$ the domains of $X_A$ and $X_B$ of different types can be considered independent form each other. Therefore, in addition to the previous symmetries,  $C$ is valid.\\
\\
$S3=\{R,C \}$\\
$\mathcal{S}_{S3}=\{ (X_A,X_B;\ell), (\hat{X}_B,\hat{X}_A;\ell), (X_B,X_A;\ell), (\hat{X}_A,\hat{X}_B;\ell), \\  (X_A,\hat{X}_B;\ell), (\hat{X}_A,X_B;\ell), (X_B,\hat{X}_B;\ell), (\hat{X}_B,X_A;\ell) \}$.

Note that $\mathcal{S}_{S1} \subset \mathcal{S}_{S2} \subset \mathcal{S}_{S3}$.  
\\

\item[-]($\ell \gg L_{M} $): $P(X_A,X_B,\ell)$ is dominated by $X_A$ and $X_B$ in different macro-structures. Note that the frequency of $X_A$ in one macro-structure and $\hat{X}_A$ in a different macro-structure are,  in general, different. Therefore,  for generic $X_A,X_B$ we have $P(X_A,X_B;\ell) \neq P(\hat{X}_B,\hat{X}_A;\ell)$, meaning that $S1$ (and thus $S2$ and $S3$) is no longer valid. On the other hand, our conjectured Chargaff symmetry, Eq.~(\ref{eq.extended}), is valid for both $X_A$ and $X_B$ separately (because they are small scale observables). Therefore $X_A$ and $X_B$  can be interchanged in the composite observable $Y$.
\\
$S4=\{RCR,C \}$\\
$\mathcal{S}_{S4}=\{ (X_A,X_B; \ell) , (\hat{X}_A,{X_B};\ell), ({X_A},\hat{X}_B;\ell), (\hat{X}_A,\hat{X}_B;\ell) \}$.

Note that $\mathcal{S}_{S4} \subset \mathcal{S}_{S3}$.

\end{itemize}

\section*{Discussion}
The complement symmetry in {\it double}-strand genetic sequences, known as the First Chargaff Parity Rule, is nowadays a trivial consequence of the double-helix assembly of DNA. However, from a historical point of view, the symmetry was one of the key ingredients leading to the double-helix solution of the complicated genetic structure puzzle, demonstrating the fruitfulness of a unified study of symmetry and structure in genetic sequences. In a similar fashion, here we show empirical evidence for the existence of new symmetries in the DNA (Figs.~\ref{fig.1}-\ref{fig.4}) and we explain these observations using a simple domain model whose key features are dictated by the role of  transposable elements in shaping DNA. In view of our model, our empirical results can be interpreted as a consequence of the action of transposable elements that generate a skeleton of symmetric domains in DNA sequences.
Since domain models are known to explain also much of the structure observed in genetic sequences, our results show that structural complex organisation of single-strand genetic sequences and their nested hierarchy of symmetries are  manifestations of the same biological processes. We expect that future unified investigations of these two features will shed light into their (up to now not completely clarified)  evolutionary and functional role.  For this aim, it is crucial to extend the analyses presented here to organisms of different complexity \cite{KFCHZZL09}. In parallel, we speculate that  the  unraveled hierarchy of symmetry at different scales could play a role in understanding how chromatin is spatially organised, related to the puzzling  functional role of long-range correlations~\cite{BWDSARSF08, AVAAdT11}.  

\section*{Methods}
\subsection*{Algorithm used to generate the synthetic sequence}

We create synthetic genetic sequences through the following implementation of the three steps of the model we proposed above:

\begin{itemize}
  
\item[(1)] The processes $p$ we use to generate genetic sequences are Markov processes of order one such that the nucleotide $s_i$ at position $i$ is drawn from a probability $P(s_i | s_{i-1}) = M_{ s_{i-1},s_i}$, where $M$ is a $4$ by $4$ stochastic matrix. The matrices $M$ are chosen such that the processes' invariant measures $\mu$ do not satisfy the Chargaff property: $\mu(A)\neq \mu(T)$ and $\mu(C)\neq \mu(G)$. The exponential decay of correlations of the Markov chains determines the domain sizes $L_D$ (in our case  $L_D\simeq 10$).

\item[(2)]  We use the processes $p$ to generate chunks of average size $150$ (the length of each chunck was drawn uniformly in the range $[130,170]$. With probability $1/2$, we applied the reverse-complement (CRC) operation to the chunck before concatenating it to the previous chunck (process $\hat{p}$). This choice implies that the typical cluster size is $L_S\simeq 2*150=300$. The process of concatenating chunks together is repeated to form a macrostructure of length $L_M \simeq 10^6$.

\item[(3)] We concatenate two different macrostructures, obtained from steps (1) and (2) with two different matrices $M_I$ and $M_{II}$:
  $$
  M_I = \begin{bmatrix}
     0.2 & 0.1 & 0.2 & 0.5 \\ 0.01 & 0.84 & 0.01 & 0.14 \\ 0.4 & 0.1 & 0.4 & 0.1 \\ 0.3 & 0.15 & 0.25 & 0.3 
    \end{bmatrix},\;     M_{II} = \begin{bmatrix}
       0.1 & 0.2& 0.1& 0.6\\ 0.1& 0.75& 0.1& 0.05\\ 0.1& 0.4& 0.1& 0.4\\ 0.1& 0.35& 0.45& 0.1
    \end{bmatrix}
    $$
where columns (and rows) corresponds to the following order:  $[A,C,G,T]$.
\end{itemize}
  
\subsection*{Data handling}
Genetic sequences of Homo Sapiens  were downloaded from the National Center for Biotechnology Information  $( ftp://ftp.ncbi.nih.gov/genomes/H\_sapiens )$. We used reference assembly build 38.2.
The sequences were processed to remove  all letters different from $A,C,G,T$ (they account for $\approx 1.66\%$ of the full genome and thus their removal has no significant impact on our results).

\subsection*{Codes}

Ref.~\cite{zenodo} contains data and codes that reproduce the figures of the manuscript for different choices of observables and chromosomes.


\section*{Acknowledgements}

EGA and GC thank the Max Planck Institute for the Physics of Complex Systems in Dresden (Germany) for hospitality and support at the early stages of this project.

\section*{Author contributions statement}
G.C.  initiated and designed the study.   E.G.A.  and G.C. contributed to the development of the study, carried out statistical analysis and wrote the manuscript. E.G.A , G.C.  and M.D.E. discussed  and interpreted the results.   E.G.A , G.C.  and M.D.E  reviewed the manuscript.

\section*{Additional information}

 \textbf{Competing interests}  The authors declare that they have no competing interests. 

\newpage
\section*{Supplementary Information}
\subsection*{Derivation of the nested hierarchy of symmetries}\label{sec.hierarchy}
We  derive  the nested hierarchy of symmetries in the minimal model for genetic sequence.

\subsubsection*{Notation and model properties}

To fix notations, we describe our model and its statistical properties as follows:
\begin{itemize}
\item The full sequence is build  concatenating  $r$ macro-structures: $\textbf{s}=\textbf{m}_1\, \textbf{m}_2 \cdots \textbf{m}_r$. 
\item A macrostructure $\textbf{m}$ is build concatenating $m$ domains: $\textbf{m}=\textbf{d}^{\sm}_1\, \textbf{d}^{\sm}_2 \cdots \textbf{d}^{\sm}_{m}$.
\item The average domain length
is denoted by $L_D$, the average macro-structure length is denoted by $L_M$. The total length of the sequence is $N$.
\item  A domain $ \textbf{d}^{\sm}$ in the macro-structure $\textbf{m}$ is a  finite-size realisation of a process chosen between  two\footnote{Generalisations to more than two symmetrically domain-types is straightforward and it is not expected to change the main features of the model.}   symmetrically related  process-types: $C_{\sm}$ and  $\hat{C}_{\sm}$. We use the notation $\textbf{d} \in C$ to indicate that $\textbf{d}$ is generated by the process of type $C$.
\item For a a given observable X, we denote by $f_C(X)$ the limiting relative frequency\footnote{We assume that process types $C$ are such that $f_C(X)$  are well defined for all choice of observables $X$ in the limit of size of domains going to infinity.} of occurrence of $X$ in  a domain  of type $C$.  Recall that, the definition of  symmetrically related processes  (of the same macro-structure)  imposes that, for every choice of $X$:
\begin{equation}\label{symProbDomains}  
f_{C}(X)=f_{\hat{C}}(\hat{X})
\end{equation}
In principle,  different macro-structures have different process-types statistics.  
 \item
 We denote by $(\underline{c},\underline{l})$ an ordered sequence of domains of types $\underline{c}:=(c_1,\cdots,c_k); \quad c_j  \in  \{ C,\hat{C}\}$ of lengths $\underline{l}:=(l_1,\cdots,l_k)$ respectively; and by $\hat{(\underline{c},\underline{l})}$ 
  the sequence of domains defined by $(\hat{c}_k,\cdots,\hat{c}_1)$ and $(l_k,\cdots,l_1)$. We denote by $\pi^{\sm}[(\underline{c},\underline{l})]$ the relative frequency\footnote{We assume that the structural properties of a given macro-structure is such that $\pi$ is well defined in the limit of number of domains $m$ going to infinity.} of  counts of subsequence of domains $(\underline{c},\underline{l})$ in $\textbf{m}$. 
\\

We denote by $\pi^{\sm}(\{c,l\})$  the relative frequency of a cluster of length $l$ of domains of the same type $c$.
\\
 For $\alpha,\beta\in  \{ C,\hat{C}\}$ and $k\ge 1$ , we denote  $\pi^{\sm}(\alpha,\beta;k)$ 
 the relative frequency of  $j$ such that  $\textbf{d}_j \in \alpha$ and $\textbf{d}_{j+k} \in \beta$. 
\\

\item In each macro-structure, the probability distribution  of domain-sizes  is denoted  $p_{\sm}(l)$. 

 \item  We do not enforce any prescription to concatenate domains in a macrostructure (determined  by $\pi$), but the following properties:
  \begin{itemize}

 \item  $\pi^{\sm}[(\underline{c},\underline{l})]=\pi^{\sm}[\hat{(\underline{c},\underline{l})}]$ This ensure that  the  structural statistics of  two symmetrically coupled domain-types ordering is unbiased.

 \item for $k>>L_S/L_D$ ; $\pi^{\sm}(c_1,c_2,k)=\pi^{\sm}(c_1)\pi^{\sm}(c_2)$. This defines the average length $L_S$ beyond which   correlations in domain ordering can be neglected. $L_S $ is  thus the average size of clusters of domains of the same type.
 \end{itemize} 

\end{itemize}

\subsubsection*{Derivation of symmetries}

We start by showing the validity of the extended Chargaff symmetry $P(X) = P(\hat{X})$ for $\ell < L_M$. We denote by $\#_{ (\underline{c},\underline{l})}(X) $ the counts of $X$ inside  $(\underline{c},\underline{l})$. Using $\pi[(\underline{c},\underline{l})] = \pi[\hat{(\underline{c},\underline{l})}]$ and $f_{C}(X)=f_{\hat{C}}(\hat{X})$  we have that $\#_{ (\underline{c},\underline{l})}(X) = \#_{ \hat{(\underline{c},\underline{l})}}(\hat{X}) $. Finally, for  $X$ of size $\ell_X \ll L_M$, the counts of $X$ in the full sequence is dominated by $X$ not overlapping different macro-structures and thus we conclude that ($N'=N-\ell$)
\begin{eqnarray}\label{first}
P(X) &\simeq& 1/N'\sum_{m} \sum_{(\underline{c},\underline{l})_m} \#_{ (\underline{c},\underline{l})}(X) \nonumber\\
&= &1/N' \sum_{m} \sum_{\hat{(\underline{c},\underline{l})}_m} \#_{ \hat{(\underline{c},\underline{l})}}(\hat{X})\nonumber\\
&\simeq& P(\hat{X}).
\end{eqnarray}

We now show the validity of the nested hierarchy of symmetries discussed in the main paper. We focus on observables of the form $Y=(X_A,X_B;\ell)$, where $X_A$ and $X_B$ are oligonucleotide of size much smaller than typical domain sizes $L_{D}$.
We always approximate the counts of $X $ inside a domain of type $C$ and of length $l$ by $l \cdot f_C(X)$.
\\

Define
\begin{eqnarray}
\#_{(i)}(Y)&:=&  \textrm{  number of $Y$ fully inside the $i$-th domain}\nonumber \\
\#_{(ij)}(X_A,X_B,\ell)&:=& \textrm{  number of $X_A$ fully in the $i$-th and $X_B$ in the $j$-th domains, at distance $\ell$} \nonumber \\
\#(Y)&:=& \textrm{  number of $Y:=(X_A,X_B,\ell)$ in the full string} \label{countsY}\nonumber\\
&=& \sum_{i} \#_{(i)}(Y)+ \sum_{i}\sum_{j>i}{\#_{(ij)}(X_A,X_B,\ell)} + \nonumber \\ &\quad&+\{ \textrm{terms where $X_A$ or $X_B$ overlap domains boundaries}\}\nonumber
\end{eqnarray}
\vspace{0.4cm}
As we will consider only the case $l_{X_A},l_{X_B}  \ll L_D$,  we neglect the last term.

We can now investigate and rule out the main contributions to the overall counting $\#(Y)$  at different scales:
\begin{itemize}
\item [-]\textbf{$(\ell \ll L_D)$}: At these scales the following sum dominates,
\begin{eqnarray*}
\#(Y) \simeq \sum_{i=1} \#_{(i)}(Y)&\simeq&
     \sum_{m=1}^r g_m(\ell)  \left[ f_{C_\sm}(Y) + f_{\hat{C}_\sm}(Y) \right] \\
    &=&   \sum_{m=1}^r g_m(\ell) \left[  f_{C_\sm}(Y) +  f_{C_\sm}(\hat{Y}) \right] 
\end{eqnarray*}
where 
\begin{eqnarray}
g_m(\ell):=\frac12\sum_{l=\ell}^{\infty} p_m(l) (l-\ell) \quad \quad \quad \ell \ll L_D.\nonumber
\end{eqnarray} 
\\
We conclude that  $\#(X_A,X_B,\ell) \simeq \#(\hat{X}_B,\hat{X}_A,\ell)$ at these scales, 
and thus symmetry $S1$ is valid.  This can also be derived directly from equation~(\ref{first}).
\vspace{1cm}
\end{itemize}
For $\ell >>L_D$, $X_A$ and $X_B$ typically lie in different domains and therefore the  second term  in equation~(\ref{countsY})  dominates
$$ \#(Y) \simeq  \sum_{i=1}\sum_{j>i}{\#_{(ij)}(X_A,X_B,\ell)}. $$
The counts will be estimated as the product of the probabilities of $X_A$ and $X_B$ because each domain is an independent realisations. At different scales $\ell$ there are different relationships between the domains in which $X_A$ and $X_B$ typically lie, leading to the following  cases:
\begin{itemize}
\item[-] \textbf{$( L_D < <\ell < L_S)$}:
At these scales the sum is dominated by counts of  $Y$ inside a  cluster of  domains  of the same type. Each cluster contribute to the counts of $Y$ with a term $\pi[\{c,l\}] (l-\ell) f_C(X_A)f_{C}(X_B)$ 
and thus, in this case
\begin{eqnarray}
\sum_{i=1}\sum_{j>i}{\#_{(ij)}(X_A,X_B,\ell)}
 &\simeq &  \sum_{m=1}^r h_m(\ell) \left[ f_{C_\sm}(X_A) f_{C_\sm}(X_B)  +  f_{\hat{C}_\sm}(X_A)f_{\hat{C}_\sm}(X_B) \right]\nonumber \\ 
&=&\sum_{m=1}^{r} h_m(\ell)   \left[ f_{C_\sm}(X_A) f_{C_\sm}(X_B)  +   f_{C_\sm}(\hat{X}_A)f_{C_\sm}(\hat{X}_B) \right] \nonumber
\end{eqnarray}
where 
\begin{eqnarray}
h_m(\ell):=\frac12\sum_{l=\ell}^{\infty} \pi[\{c,l\}]  (l-\ell) \quad \quad \quad L_D < \ell \ll L_S \nonumber
\end{eqnarray} 
\\
We conclude that  $\#(X_A,X_B,\ell) \simeq \#(\hat{X}_B,\hat{X}_A,\ell)\simeq \#(X_B,X_A,\ell) \simeq \#(\hat{X}_A,\hat{X}_B,\ell)$ at these scales, 
and thus symmetry $S2$ (and $S1$) is valid. If the processes are such that correlations inside domains vanishes at a scale smaller than the realization of the process, we consider this shorter correlation time to be the effective domain size $L_D$ and $S2$ sets in at this shorter scale.
\\
\\
\item[-]$( L_S \ll \ell \ll L_M)$: At these scales the sum is dominated by $X_A$ and $X_B$ lying in different cluster
\begin{eqnarray}
\sum_{i}\sum_{j>i}{\#_{(ij)}(X_A,X_B,\ell)}
  &= &  s(\ell) \sum_{m=1}^r  \left[ f_{C_\sm}(X_A) f_{C_\sm}(X_B)+f_{C_\sm}(X_A) f_{\hat{C}_\sm}({X}_B) + \right. \nonumber\\
  && \left. \quad \quad \quad \quad \quad+ f_{\hat{C}_\sm}({X}_A) f_{C_\sm}(X_B) + f_{\hat{C}_\sm}({X}_A)f_{\hat{C}_\sm}({X}_B) \right]\nonumber\\
  &= &  s(\ell) \sum_{m=1}^r  \left[ f_{C_\sm}(X_A) f_{C_\sm}(X_B)+f_{C_\sm}(X_A) f_{{C}_\sm}(\hat{X}_B)+\right. \nonumber\\
  && \left. \quad \quad \quad \quad \quad+  f_{{C}_\sm}(\hat{X}_A) f_{C_\sm}(X_B) + f_{{C}_\sm}(\hat{X}_A)f_{{C}_\sm}(\hat{X}_B) \right]\nonumber\\
   &= &  s(\ell) \sum_{m=1}^r  \left[ \left( f_{C_\sm}(X_A)+  f_{{C}_\sm}(\hat{X}_A) \right) \left(f_{C_\sm}(X_B)+ f_{{C}_\sm}(\hat{X}_B) \right)\right]\nonumber\\
\end{eqnarray}
where  
\\
\begin{eqnarray}
s(\ell)\simeq \frac14(L_M-\ell)  \quad \quad \quad L_S<\ell\ll L_M.\nonumber
\end{eqnarray} 
\\
We conclude that  $\#(X_A,X_B,\ell) \simeq \#(\hat{X}_B,\hat{X}_A,\ell)\simeq \#(X_B,X_A,\ell) \simeq \#(\hat{X}_A,\hat{X}_B,\ell)\simeq
\#(\hat{X}_A,X_B,\ell) \simeq \#(\hat{X}_B,{X}_A,\ell)\simeq \#(X_B,\hat{X}_A,\ell) \simeq \#({X}_A,\hat{X}_B,\ell)$ at these scales, 
and thus symmetry $S3$ (and $S2,S1$ and $S4$) is valid.

\item[-] \textbf{$(L_M \ll \ell) $}: At these scales the sum is dominated by counts where $X_A$ and $X_B$  are in different macro-structures:
\begin{eqnarray}
\sum_{i}\sum_{j>i}{\#_{(ij)}(X_A,X_B,\ell)}
      &= & \sum_{m=1}^r\sum_{n>m} q_{m,n}(\ell)\left[ \left( f_{C_\sm}(X_A)+  f_{{\hat{C}}_\sm}({X}_A) \right) \left(f_{C_\sn}(X_B)+ f_{\hat{C}_\sn}({X}_B) \right)\right] \nonumber\\
   &= & \sum_{m=1}^r\sum_{n>m} q_{m,n}(\ell) \left[ \left( f_{C_\sm}(X_A)+  f_{{C}_\sm}(\hat{X}_A) \right) \left(f_{C_\sn}(X_B)+ f_{{C}_\sn}(\hat{X}_B) \right)\right].\nonumber
\end{eqnarray}

where $q_{m,n}(\ell)$ counts how many sites separated by $\ell$ lie in macro-structures $\textrm{m}$ and  $\textrm{n}$, respectively. \\
We conclude that  $\#(X_A,X_B,\ell) \simeq \#(\hat{X}_A,X_B,\ell)\simeq \#(X_A,\hat{X}_B,\ell) \simeq \#(\hat{X}_A,\hat{X}_B,\ell)$ and thus symmetry $S4$ is valid.
\end{itemize}

\end{document}